\documentclass[11pt]{amsproc}
\usepackage{mathrsfs, color}
\usepackage{stmaryrd}
\usepackage{cases}
\usepackage{amssymb}
\usepackage{amsmath}
\usepackage{amsfonts}
\usepackage{graphicx}
\usepackage{amsmath,amstext,amsbsy,amssymb}

\renewcommand{\a}{\alpha}

\newcommand{\ra}{\rangle}
\newcommand{\lan}{\langle}

\def\Tr{\mathrm{Tr}}

\begin{document}


\title[Improved unitary uncertainty relations]
{Improved unitary uncertainty relations}

\author{Xiaoli Hu}
\address{School of Artificial Intelligence, Jianghan University, Wuhan, Hubei 430056, China}
\email{xiaolihumath@163.com (co-corresponding author)}

\author{Naihuan Jing}
\address{Department of Mathematics, North Carolina State University, Raleigh, NC 27695, USA}
\email{jing@ncsu.edu (co-corresponding author)}

\begin{abstract}
We derive strong variance-based uncertainty relations for arbitrary two and more unitary operators
by re-examining the mathematical foundation of the uncertainty relation. This is achieved
by strengthening the celebrated Cauchy-Schwarz inequality using a method of brackets and convex functions.
The unitary uncertainty relations outperform several strong unitary
uncertainty relations, notably
 better than some recent best lower bounds such as [Phys. Rev. Lett. 120, 230402 (2018)] and [Phys. Rev. A. 100, 022116 (2019)].
\end{abstract}


\maketitle
\section{Introduction}

The uncertainty relation lies at the foundation of the quantum theory with
a wide range of applications in quantum information including quantum cryptography \cite{1,2}, quantum entanglement \cite{3,4,5}, quantum speed limit \cite{6}, signal processing
\cite{7}, and so on. Uncertainty relations can be tested experimentally with neutronic \cite{8,9} and photonic qubits \cite{10,11,12,13}. There are two kinds of uncertainty relations in quantum theory: the preparation uncertainty relation and the measurement uncertainty relation.

The well-known Heisenberg uncertainty principle \cite{15} has played a pivotal
role and provided deep insights into the nature of quantum world and distinguished itself from the classical world.
The classical Heisenberg uncertainty relation says that
the measurement of position $x$ and momentum $p$ cannot be done exactly and simultaneously
\begin{equation}\label{eq:Heisen 1}\Delta x\Delta p\geq 1/2,
\end{equation}
where $[x, p] = iI$ ($I$ is the identity operator)
and $\Delta$ is the variance. The modern form of the uncertainty principle was established by
Kennard \cite{16} and well formulated by Weyl \cite{17}.
In 1929, Robertson generalized \eqref{eq:Heisen 1} to an arbitrary pair of non-commuting bounded
observables $A$ and $B$ and showed that the product of variances of $A$ and $B$  has a general lower bound \cite{18}:
\begin{equation}\label{eq:Rober 2}\Delta A\Delta B\geq |\lan [A,B]\ra|/2.
\end{equation}
where $\lan [A,B]\ra$ is the expectation value of the commutator $[A, B]$ with respect to a fixed state.
Further works on variance-based uncertainty relations have been carried out in \cite{19, 20}.
One notices that \eqref{eq:Heisen 1} or \eqref{eq:Rober 2} maybe trivial and
do not capture the intrinsic incompatibility of non-commuting observables \cite{19, 20},
this triggered further formulation of the uncertainty relation.
One of the most promising ones is perhaps the
entropy-based uncertainty relation \cite{21,22,23,24,25,26,XJ2}.

Another important direction is to improve the uncertainty relation based on variances in a nontrivial way.
In 2014, Mcconne and Pati \cite{MP} proposed a sum form of variance-based uncertainty relation, which captures
the essence of noncommutativity and also raises the question of compatibility (see \cite{XJ1} for further
discussion). In \cite{BP} the sum form was strengthened considerably. In \cite{Busch} the authors showed that there are no nontrivial unconditional joint-measurement bounds for state-dependent errors in the conceptual framework, while Heisenberg-type measurement uncertainty relations for state-independent errors have been proven (cf. \cite{Xiao2019, Ozawa2003}).
Also the uncertainty relation for the Fourier transform has been studied in \cite{SP}.
In \cite{Rud}, the authors presented the uncertainty relation for the characteristic functions of the quantum mechanical position and momentum probability distributions. Yet another important development is to improve the generalized product form of variance-based uncertainty relation given by Bong et al \cite{Bong},
which descends to Heisenber-Robertson's uncertainty relation for two unitary operators
\cite{H2, R2, JSL} at a special situation. 
Bong's beautiful bound is expressed implicitly in terms of the Gram determinant,
which is sometimes hard to extract the exact form. In \cite{Yu} one of us and collaborators
have proposed an improvement to evaluate the lower bound using a sequence of ``fine-grained''
inequalities \cite{XJ3}.

The goal of this paper is to give another nontrivial improvement on the uncertainty relation for general two and more
unitary operators.
We achieve this by examining the mathematical foundation of the uncertainty relation to improve the Cauchy-Schwarz inequality.
We use brackets and convex functions to derive variance-based unitary uncertainty relations in the products form for two unitary operators in all quantum systems, which
is simple to understand and can be well presented. We show that our new bound outperforms
Bong et al.'s \cite{Bong} in the whole range as well as Yu et al.'s \cite{Yu} and Li et al.'s \cite{JSL} in almost the whole interval.

This paper is organized as follows. In section 2 we introduce the method of brackets and convex functions
based on the Cauchy-Schwarz inequality to derive
our main results, which gives effective lower bounds for the product of variances.
In section 3 we examine several examples to show that our new bounds are indeed tighter than those of Bong et al's and Yu et al's
bounds. In section 4 we discuss how to generalize our results to multiple observables. The conclusion is given in section 5.

\section{UUR for two unitary operators}
Let $|\psi\ra$ be a fixed quantum state in a Hilbert space $H$ and
$A$ a unitary operator on $H$. Let $\lan A\ra=\lan\psi | A|\psi\ra$ be the expectation value of $A$. 
Let $\delta A=A-\lan A\ra$, then the variance of $A$ is defined by $\Delta A^2=\lan(\delta A)^{\dagger}(\delta A)\ra$.
The celebrated unitary uncertainty relation (UUR) says that
\begin{equation}\label{e:uur}
\Delta A^2\Delta B^2\geq|\lan A^{\dagger}B\ra-\lan A^{\dagger}\ra\lan B\ra|^2.
\end{equation}
This fundamental relation is vital to quantum mechanics. Usually the UUR follows
from the well-known Cauchy-Schwarz inequality, which we now recall to motivate our discussion.

Fix a computational basis $\{|j\ra\}_1^n$ of $H$, and let $\a=(\a_1,\cdots,\a_n)$ be the coordinate vector of $A$,
and similarly $\beta$ for $B$.
Physically the quantities $|\alpha_i|=x_i$
and $|\beta_i|=y_i$ are observed real numbers. Let
$|x|=\sqrt{x\cdot x}$ be the vector norm. Then
\begin{equation}\label{eqr1}
\begin{split}
\Delta A^2\Delta B^2&=\lan\psi|(\delta A)^{\dagger}(\delta A)|\psi\ra\cdot\lan\psi|(\delta B)^{\dagger}(\delta B)|\psi\ra\\
&=\sum_{i=1}^n|\a_i|^2\cdot \sum_{i=1}^n|\beta_i|^2
=|x|^2|y|^2,
\end{split}
\end{equation}
\begin{equation}\label{eql2}
\begin{split}
|\lan A^{\dagger}B\ra-\lan A^{\dagger}\ra\lan B\ra|^2&=|\lan\psi|(\delta A)^{\dagger}(\delta B)|\psi\ra|^2=|\sum_{i=1}^n\a^*_i\beta_i|^2
=(x\cdot y)^2.
\end{split}
\end{equation}
The Cauchy-Schwarz inequality $|x\cdot y|\leq |x||y|$ immediately implies the UUR \eqref{e:uur}
\cite{Lax}.
Therefore an improvement of the Cauchy-Schwarz inequality will lead to an improvement of the UUR. This takes us to the following discussion.

Denote by $\vec{x}_m=(x_1,\cdots,x_m,0,\cdots,0)$ a partial vector of $x=(x_1, \cdots, x_n)$.
Define $\vec{x}^c_m=x-\vec{x}_m=(0, \cdots, x_{m+1}, \ldots, x_n)$.
The Cauchy-Schwarz inequality implies that
\begin{equation}\label{eq6}
\begin{split}
&(x\cdot y)^2=(\vec{x}_m\cdot \vec{y}_m)^2+(\vec{x}^c_m\cdot \vec{y}^c_m)^2+2(\vec{x}_m\cdot \vec{y}_m)(\vec{x}^c_m\cdot \vec{y}^c_m)\\
&\leq |\vec{x}_m|^2|\vec{y}_m|^2+|\vec{x}^c_m|^2|\vec{y}^c_m|^2+2|\vec{x}_m||\vec{y}_m||\vec{x}^c_m||\vec{y}^c_m|\\
&\leq |\vec{x}_m|^2|\vec{y}_m|^2+|\vec{x}^c_m|^2|\vec{y}^c_m|^2+|\vec{x}_m|^2|\vec{y}^c_m|^2+|\vec{x}^c_m|^2|\vec{y}_m|^2\\
&=(|\vec{x}_m|^2+|\vec{x}^c_m|^2)(|\vec{y}_m|^2+|\vec{y}^c_m|^2)=|x|^2|y|^2
\end{split}
\end{equation}
with equality if and only if $\vec{x}_m=k\vec{y}_m$, $\vec{x}^{c}_m=k^c\vec{y}^{c}_m$ and $|\vec{x}_m||\vec{y}_m^{c}|=|\vec{x}_m^{c}||\vec{y}_m|$, where $k$ and $k^c$ are constants.

We have the following result on UUR.

\emph{Theorem 1.} \label{t:bound}
Let $K_m=(|\vec{x}_m||\vec{y}_m|+|\vec{x}^c_m||\vec{y}^c_m|)^2$ for $m\in \{1,\cdots,n\}$
and $K_m^v=v K_m +(1-v)|x|^2|y|^2$ a convex function, then
\begin{equation}\label{e:bound1a}
\begin{split}
|\lan A^{\dagger}B\ra-\lan A^{\dagger}\ra\lan B\ra|^2&\leq K_m\leq K_m^v\leq
\Delta A^2\Delta B^2,
\end{split}
\end{equation}
where $v\in [0,1]$.

\emph{Proof}. Note that $(x\cdot y)^2\leq K_m\leq |x|^2|y|^2$ was already shown in \eqref{eq6}. Now
$(K_m^v)'=K_m-|x|^2|y|^2 \leq 0$, so $K_m^v$ is non-increasing function. Then we have
\begin{equation*}(x\cdot y)^2\leq K_m= f(1)\leq K_m^v\leq f(0)= |x|^2|y|^2.
\end{equation*}
So \eqref{e:bound1a} follows from \eqref{eqr1} and \eqref{eql2}.
\hskip\fill$\square$

In reality one usually quantifies $m$ coordinates of $x$ not necessarily the first $m$ ones. Physically
one can just sample any $m$ coordinates of $x$. Mathematically this is better achieved by
using symmetry.
Let $S_n$ be the symmetric group acting naturally
on $\mathbb R^n$ by permutation.
 In particular
$\sigma(\vec{x}_m)=(x_{\sigma(1)},\cdots, x_{\sigma(m)},0,\cdots,0)$, then
$\sigma(\vec{x}_m^c)=\sigma(x)-\sigma(\vec{x}_m)$.
We set
\begin{equation}\label{e:Km}
\sigma(K_m)=(|\sigma(\vec{x}_m)||\sigma(\vec{y}_m)|+|\sigma(\vec{x}^c_m)||\sigma(\vec{y}^c_m)|)^2,
\end{equation}
 then we have the second result on UUR.

\emph{Theorem 2.}\label{t:bound2}
Let $\tilde{K}_m=\max_{\sigma\in S_n} \{\sigma(K_m)\}$ and $\tilde{K}=\max_m\{\tilde{K}_m\}$, then we have
\begin{equation}\label{e:bound2}
\lan A^{\dagger}B\ra-\lan A^{\dagger}\ra\lan B\ra|^2\leq K_m\leq\tilde{K}_m\leq\tilde{K}\leq \Delta A^2\Delta B^2.
\end{equation}

Note that $\tilde{K}_m=\tilde{K}_{d-m}$ by \eqref{e:Km}, so we only need $\tilde{K}_m$ for $m=1, \cdots, [\frac d2]$.
Set $\tilde{K}^v_m=v\tilde{K}_m+(1-v)|x|^2|y|^2$, then
\begin{equation}
|\lan A^{\dagger}B\ra-\lan A^{\dagger}\ra\lan B\ra|^2\leq \tilde{K}_m\leq \tilde{K}^v_m\leq
\Delta A^2\Delta B^2.
\end{equation}


\section{Pure or Mixed, that is a Question}

In this section, we are interested in the question why we have focused only on the uncertainty associated with pure states
from perspectives of both physics and mathematics. Previous works on variance-based uncertainty relations have considered quantifying uncertainties in terms of both pure and mixed states. Here, in contrast, we would like to show that the study on variance-based uncertainty relations for mixed states are superfluous.

Physically, the uncertainty principle exhibits a fundamental limit to the precisions with respect to multiple incompatible measurements, such as position and momentum, or measurements with mutually unbiased bases, just to name a few. More precisely, Heisenberg's uncertainty principle says that more accurately we know from one of the incompatible measurements, less accurately we know the other. However, for mixed states, we can hardly know any observable with absolute precision from measurements, thus the uncertainty relation for mixed states, which includes the variance-based formalism, goes against the original idea of Heisenberg's uncertainty principle.

Mathematically, the variance-based uncertainty relations can be reformatted into the following nonlinear optimization problem:
\begin{align}
\min \quad &
\Tr[ ( A - \langle A \rangle )^{2} \rho]~
\Tr[ ( B - \langle B \rangle )^{2} \rho]
\notag\\
\text{s.t.} \quad &
\Tr[\rho]=1, \quad \rho \geqslant 0.
\end{align}
In general one cannot expect to solve above problem effectively. Fortunately, here we are only interested in whether the minimum is achieved by pure states or mixed states. Given a quantum state $\rho$ and its spectral decomposition as
\begin{align}
\rho
=
\sum_{j} \lambda_{j} u_{j},
\end{align}
where $u_{j}:=|u_{j}\rangle\langle u_{j}|$. To avoid ambiguity, in this section we denote $\Tr[ ( A - \langle A \rangle )^{2} \rho]$ as $\Delta A^{2}_{\rho}$, and thus we have $\Delta A^{2}_{u_{j}} = \langle u_{j}| ( A - \langle A \rangle )^{2} |u_{j}\rangle$. As a consequence of the convexity of the quadratic function, we immediately obtain
\begin{align}
\Delta A^{2}_{\rho}
\geqslant
\sum_{j} \lambda_{j} \Delta A^{2}_{u_{j}},
\end{align}
which further implies that
\begin{align}
\Delta A^{2}_{\rho}\Delta B^{2}_{\rho}
&\geqslant
\left(\sum_{j} \lambda_{j} \Delta A^{2}_{u_{j}}\right)
\left(\sum_{k} \lambda_{k} \Delta B^{2}_{u_{k}}\right)\notag\\
&=
\sum_{jk}  \lambda_{j} \lambda_{k}
\Delta A^{2}_{u_{j}} \Delta B^{2}_{u_{k}}\notag\\
&\geqslant
\sum_{j} \lambda_{j}^{2}
\Delta A^{2}_{u_{j}} \Delta B^{2}_{u_{j}}.
\end{align}
Finally, by setting $u \in \{u_{j}\}_{j}$ as the pure state that minimizes $\Delta A^{2}_{u_{j}} \Delta B^{2}_{u_{j}}$ for all $j$, we get
\begin{align}\label{product pure}
\Delta A^{2}_{\rho}\Delta B^{2}_{\rho}
\geqslant
\Delta A^{2}_{u} \Delta B^{2}_{u}.
\end{align}
In other words, for any quantum state $\rho$, we can always find a pure state $u$ that minimizes the variance-based uncertainty relation. Observe that the statement is not only true for variance-based uncertainty relations with product form, but also works for variance-based uncertainty relations with sum form \cite{Xiao2019}, whose proof is given by the following inequalities
\begin{align}
\Delta A^{2}_{\rho} + \Delta B^{2}_{\rho}
&\geqslant
\sum_{j} \lambda_{j} \left( \Delta A^{2}_{u_{j}} +
\Delta B^{2}_{u_{j}}\right)\notag\\
&\geqslant
\Delta A^{2}_{v} +\Delta B^{2}_{v},
\end{align}
where $v \in \{u_{j}\}_{j}$ is the pure state that minimizes $\Delta A^{2}_{u_{j}} + \Delta B^{2}_{u_{j}}$ for all $j$. Now we conclude that, for the variance-based uncertainty relations, we only need to consider the case of pure states, namely

\emph{Lemma 1.}[Pure State Lemma]
\label{pure}
If $f(A,B,\rho)$ is the uncertainty relations, expressed as $\Delta A^{2}_{\rho}\Delta B^{2}_{\rho}$ or $\Delta A^{2}_{\rho} + \Delta B^{2}_{\rho}$, for a quantum state $\rho \in \mathcal{D}$ with $\mathcal{D}$ stands for the collection of all quantum states, then $\min_{\rho\in\mathcal{D}}f(A,B,\rho)$ is achieved by some pure states; that is
\begin{align}
\min_{\rho\in\mathcal{D}}f(A,B,\rho)
=f(A,B,\psi)
\end{align}
holds for some pure state $\psi$.

Thanks to the lemma, one only needs to investigate variance-based uncertainty relations for pure states in this work.

\section{Examples}
Let's first recall the celebrated unitary uncertainty relation (UUR) using Gram determinant. For $n+1$ unitary operators
$U_0=I, U_1, \cdots, U_n$, and $|\psi\ra$ a quantum state, the Gram matrix $G=(G_{jk})_{(n+1)\times(n+1)}$, where
$G_{jk}:=\lan U_j^{\dagger}U_k\ra$. The fact of $G$ being positive semi-definite
is formulated as a UUR for the operators $U_1, \ldots, U_n$ by Bong et al's \cite{Bong}. In particular, for two operators $A$ and $B$,
the Gram determinant
\begin{equation}
\det\begin{pmatrix} 1 & \langle A\rangle & \langle B\rangle\\
\langle A^{\dagger}\rangle & 1 & \langle A^{\dagger}B\rangle\\
\langle B^{\dagger}\rangle & \langle B^{\dagger}A\rangle & 1 \end{pmatrix}\geq 0
\end{equation}
gives the UUR as
\begin{equation}
\Delta A^2\Delta B^2\geq|\lan A^{\dagger}B\ra-\lan A^{\dagger}\ra\lan B\ra|^2.
\end{equation}

In \cite{Yu} Yu et al derived strong variance-based uncertainty relations for two unitary operators
by using a sequence of ``fined-grained'' inequalities:
\begin{equation}
I_1\geq\cdots \geq I_d\geq\cdots \geq I_n,
\end{equation}
where for $1\leq d\leq n$
\begin{equation}
I_d=\sum_{i=1}^nx_i^2y_i^2+\sum_{i<j,d<j}(x_i^2y_j^2+x_j^2y_i^2)
+\sum_{i<j\leq d}2x_iy_ix_jy_j.
\end{equation}
In this case, the result shows that $I_1=\Delta A^2\Delta B^2$, $I_n=|\lan A^{\dagger}B\ra-\lan A^{\dagger}\ra\lan B\ra|^2:=LB$ is Bong et al's lower bound, while $I_d$ provide better bounds with $I_2$ the most optimal one.

Another improved uncertainty relation based on \cite{Yu} was given by Li et al \cite{JSL}. They defined a quantity
$I_1^{'}$ which satisfies $\Delta A^2\Delta B^2\geq I_1^{'}\geq I_2$, where
$$I_1^{'}=\sum_{i=1}^nx_i^2y_i^2+\sum_{j\neq1,i\neq j}x_i^2y_j^2+y_1^2\sum_{i=4}^nx_i^2+2y_1^2x_2x_3.$$

In the following we give examples to compare performances of these three sets of bounds. We will show that our bounds are easier
 to calculate and tighter in most cases. Recall that
$$(x\cdot y)^2\leq K_m\leq K^{v}_m\leq |x|^2|y|^2, v\in [0,1].$$

\emph{Example} 1. \label{eg1} Let $H$ be an $d$-dimensional Hilbert space and $|\psi\ra=cos\theta|0\ra-sin\theta|d-1\ra\in H$ a pure state. Let $I_r$ be
the identity operator of size $r$, and consider
two unitary operators $A, B$:
\begin{equation*}A=diag(1,\omega,\omega^2,\cdots,\omega^{d-1}), \omega=e^{\frac{2\pi i}{d}}, B=\begin{bmatrix}0&1\\I_{d-1}&0\end{bmatrix}.
\end{equation*}
Note that $AB=\omega BA$. Then the associated real vectors $x=(x_1,x_2,\cdots,x_d)$ and $y=(y_1,y_2,\cdots,y_d)$ are given by
\begin{equation*}
\begin{split}
x_1=|(1-e^{-\frac{2\pi i}{d}})sin^2\theta cos\theta|,\ \
 x_2=\cdots=x_{d-1}=0,\ \
 x_d=|(1-e^{-\frac{2\pi i}{d}})sin\theta cos^2\theta|
\end{split}
\end{equation*}
and
\begin{equation*}
\begin{split}
y_1=|sin^3\theta|,y_2=|cos\theta|,\ \
y_3=\cdots=y_{d-1}=0,\ \
y_{d}=|sin^2\theta cos\theta|.
\end{split}
\end{equation*}
The lower bounds $I_i(i=1,\cdots,d)$ of the UUR \cite{Yu} for $A$ and $B$ can be computed as follows:
\begin{equation*}
\begin{split}
I_1&=|1-e^{-\frac{2\pi i}{d}}|^2\cdot|sin^6\theta cos^2\theta+sin^2\theta cos^4\theta|
=\Delta A^2\Delta B^2,\\
I_2&=|1-e^{-\frac{2\pi i}{d}}|^2\cdot|sin^6\theta cos^2\theta+sin^2\theta cos^6\theta|
=\cdots=I_{d-1},\\
I_d&=|1-e^{-\frac{2\pi i}{d}}|^2\cdot|sin^6\theta cos^2\theta|
=L_B=|\langle A^{\dagger}B\rangle-\langle A^{\dagger}\rangle\langle B\rangle|^2.
\end{split}
\end{equation*}
The lower bound $I_1^{'}$ of the UUR \cite{JSL} for $A, B$ can be calculated as follows:
\begin{equation*}
I_1^{'}=|1-e^{-\frac{2\pi i}{d}}|^2|sin^{8}\theta cos^2\theta+sin^6\theta cos^6\theta+sin^2\theta cos^4\theta|.
\end{equation*}
Our lower bounds $K_m$ and $K^{v}_m$ are given by
\begin{align*}
K_m=&|1-e^{-\frac{2\pi i}{d}}|^2|sin^{10}\theta cos^2\theta+sin^6\theta cos^6\theta+sin^4\theta cos^4\theta|\\
&\cdot2|\sqrt{sin^4\theta cos^2\theta(sin^6\theta+cos^2\theta)}
+\sqrt{sin^6\theta cos^6\theta}|^2,\\
K^{v}_m=&v K_m+(1-v)I_1.
\end{align*}

\begin{figure}[!htb]
\centering
\includegraphics[width=7.5cm,height=3.5cm]{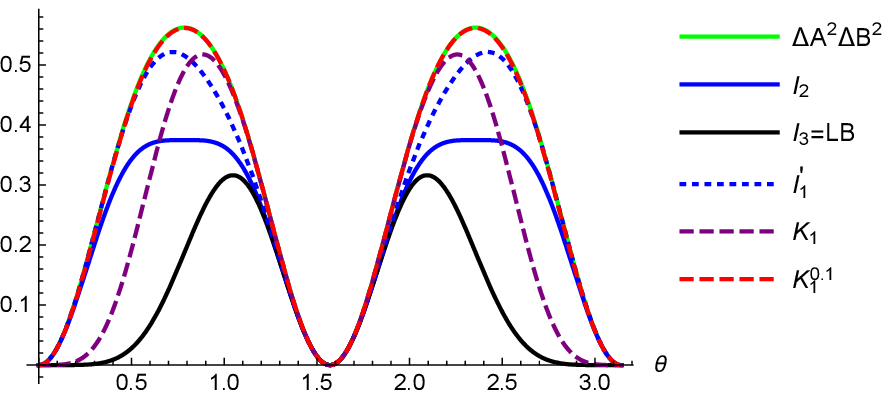}
\includegraphics[width=7.5cm,height=3.5cm]{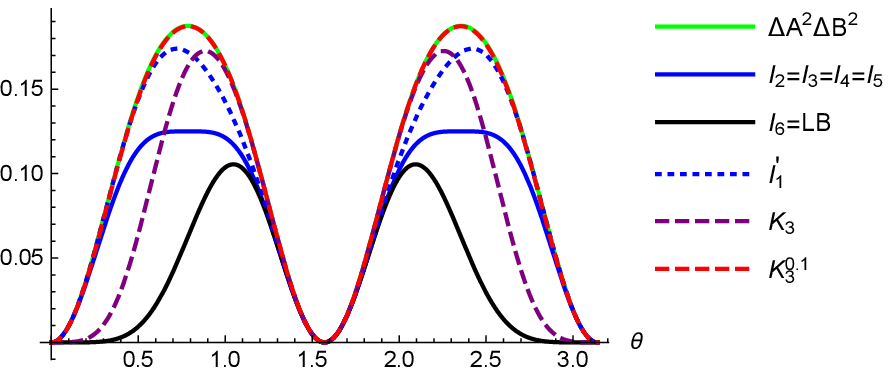}\\
d=3\ \ \ \ \ \ \ \ \ \ \ \ \ \ \ \ \ \ \ \ \ \ \ \ \ \ \ \ \ \ \ \ \  \ \ \ \ \ \ \ \ \ \ \ \  \ \ \ \ \ \
d=6\ \ \ \ \ \ \ \ \  \ \ \ \ \ \
\caption{\textbf{Comparison of bounds I.} The upper solid green, lower solid black, solid blue and dotted blue curves are bounds $\Delta A^2\Delta B^2$, Bong et al's $LB$, Yu et al's $I_2$, and Li et al's $I_1^{'}$ respectively. Our bounds $K_m$ and $K^{0.1}_m$ are shown in purple dashed and red dashed curves respectively.}\label{gra1}
\end{figure}

The case $d=2$ is trivial, as all the bounds are the same in this case:
$\Delta A^2\Delta B^2=I_1=I_2=L_B=I_1^{'}=K_m=K^{0.1}_m$.
We have plotted the bounds on the interval $[0,\pi]$ for $d=3$ and $d=6$ in Figure \ref{gra1}.
One can see that our bounds $K_1^{0.1}$ and $K_3^{0.1}$ outperform those of Bong et al, Yu et al and Li et al.

\begin{figure}[!ht]
\centering
\includegraphics[width=7cm,height=3.5cm]{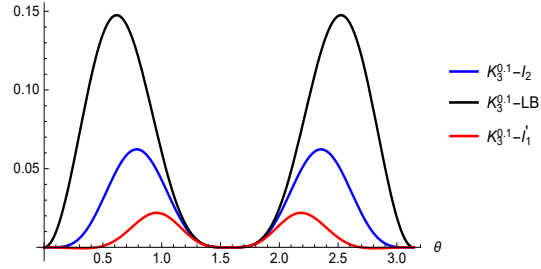}
\caption{\textbf{Bound differences I.} The solid black, blue and red curves represent $K_3^{0.1}-LB$, $K_3^{0.1}-I_2$ and $K_3^{0.1}-I_1^{'}$ respectively when $d=6$. The black curve is on the top means that our bound is the best.}\label{gra2}
\end{figure}
For $d=6$, the bound $K^{0.1}_3$ is also the best among the four bounds (see Figure \ref{gra2}).
The bounds $K_m$ and $K_m^v$ can be strengthened by Theorem 2 (\ref{t:bound2}). Take $\sigma=(36)$ for $d=6$, then
$\tilde{K}_3=\Delta A^2\Delta B^2$. For general $d$, we take $\sigma=(3~d)$, then $\tilde{K}_m=\Delta A^2\Delta B^2$.

\emph{Example} 2. Let $A$ and $B$ be the unitary operators as in Example 1. The fixed state is $|\psi\ra=\frac{1}{\sqrt{d-1}}cos\theta\sum_{k=0}^{d-2}|k\ra-sin\theta|d-1\ra$
on the $d$-dimentional Hilbert space ($d\geq 3$). Take $d=4$ for example,
\begin{equation}
A=\begin{bmatrix}1& 0&0&0\\
0&e^{\frac{i\pi}{2}}&0&0\\
0&0&e^{i\pi}&0\\
0&0&0&e^{\frac{4i\pi}{3}}\end{bmatrix},
B=\begin{bmatrix}0& 0&0&1\\
1&0&0&0\\0&1&0&0\\0&0&1&0\end{bmatrix}.
\end{equation}

\begin{figure}[!ht]
\centering
\includegraphics[width=8cm,height=4.5cm]{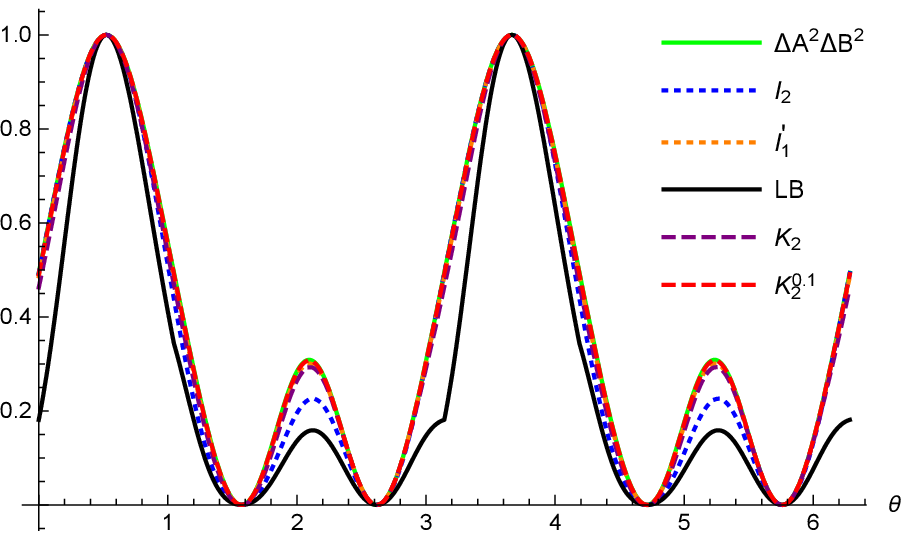}
\includegraphics[width=6cm,height=4.5cm]{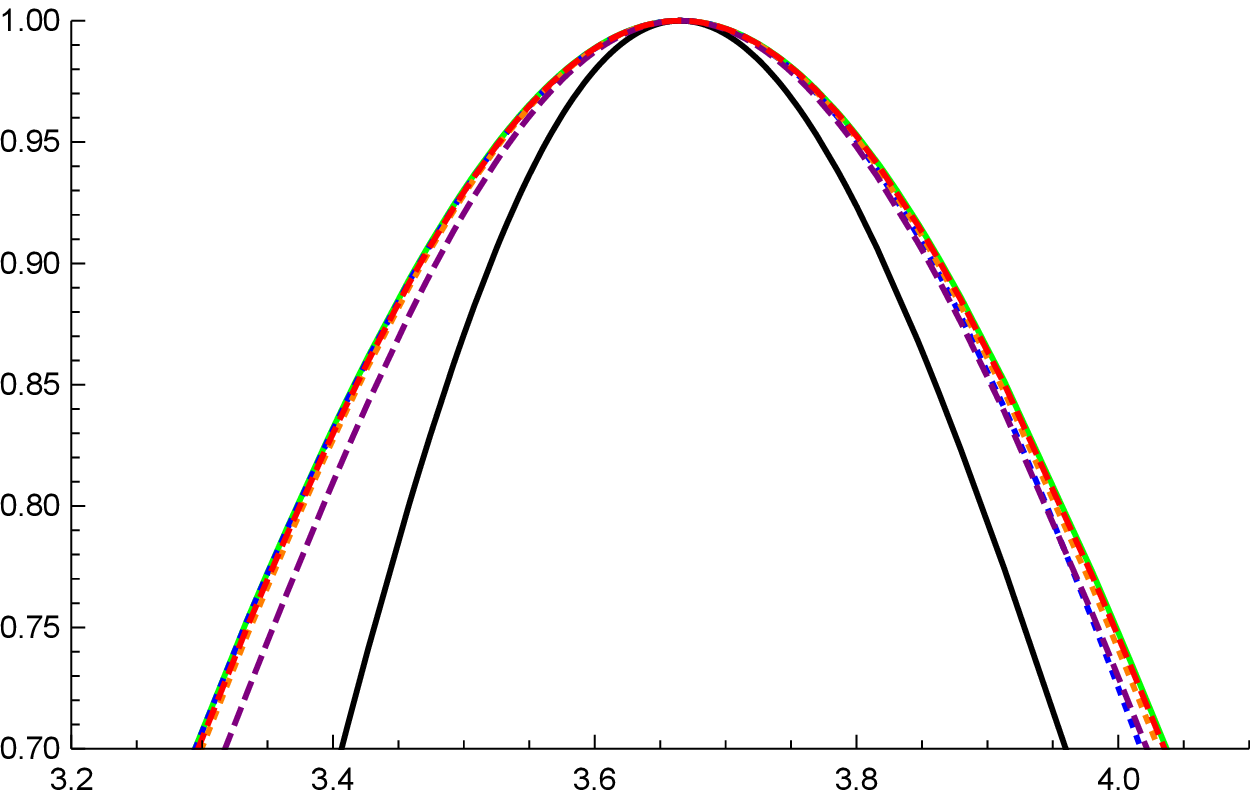}
\caption{\textbf{Comparison of bounds II.} The upper solid green, the lower black, the blue dotted and the orange dotted curves are $\Delta A^2\Delta B^2$, Bong et al's bound $LB$, Yu et al's bound $I_2$ and Li et al's bound $I_1^{'}$
respectively. Our bounds $K_2$ and $K^{0.1}_2$ are shown in purple and red dashed curves respectively.} \label{grap3}
\end{figure}

\begin{figure}[!ht]
\centering
\includegraphics[width=9cm,height=4cm]{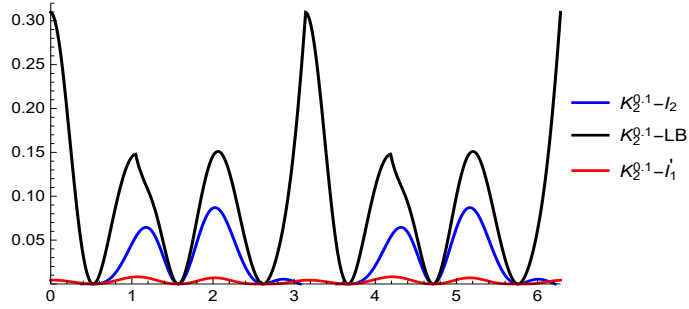}\\
\caption{\textbf{Bound differences II.} The solid black, blue and red curves represent $K_2^{0.1}-LB$, $K_2^{0.1}-I_2$ and $K_2^{0.1}-I_1^{'}$ respectively. These curves show that our bound is the tightest one.}\label{grap4}
\end{figure}

The lower bounds $K_2$ and $K_2^{0.1}$ can be calculated by Theorem 1(\ref{t:bound}). The bounds are drawn in figure \ref{grap3} and it is seen that our bound is tighter than Bong et al's, Yu et al's and Li et al's \cite{Bong, Yu, JSL}(see figure \ref{grap4}).

In this example, our bound $K^{0.1}_2$ is tighter than those of Yu et al \cite{Yu} and Bong et al \cite{Bong} in the whole range.

\emph{Example} 3.
Let us consider the pure state $|\psi\rangle=\frac{\sqrt{2}}{2}cos\theta|0\ra+\frac{\sqrt{2}}{2}cos\theta|1\ra+sin\theta|2\ra$ on a 3-dimensional Hilbert space. The unitary operators $A$ and $B$ are as in Example 1 for $d=3$:
\begin{equation*}
\begin{split}
A=\begin{bmatrix} 1&0&0\\0&e^{\frac{i\pi}{2}}&0\\0&0&e^{\frac{3i\pi}{2}}\end{bmatrix},
 B=\begin{bmatrix} 0&0&1\\1&0&0\\0&1&0 \end{bmatrix}.
  \end{split}
\end{equation*}

Using theorem 1 (\ref{t:bound}), the lower bounds can be quickly computed. Figure 5 compares these lower bound curves $LB, I_2, I_1^{'}$ with our bounds $K_2, K_2^{0.1}$. In Figure 6, their differences are shown to facilitate the comparison. It is shown that our bound $K_2^{0.1}$ is better than the previous bounds to some extent.
\begin{figure}
  \centering
  \includegraphics[width=7.5cm,height=4cm]{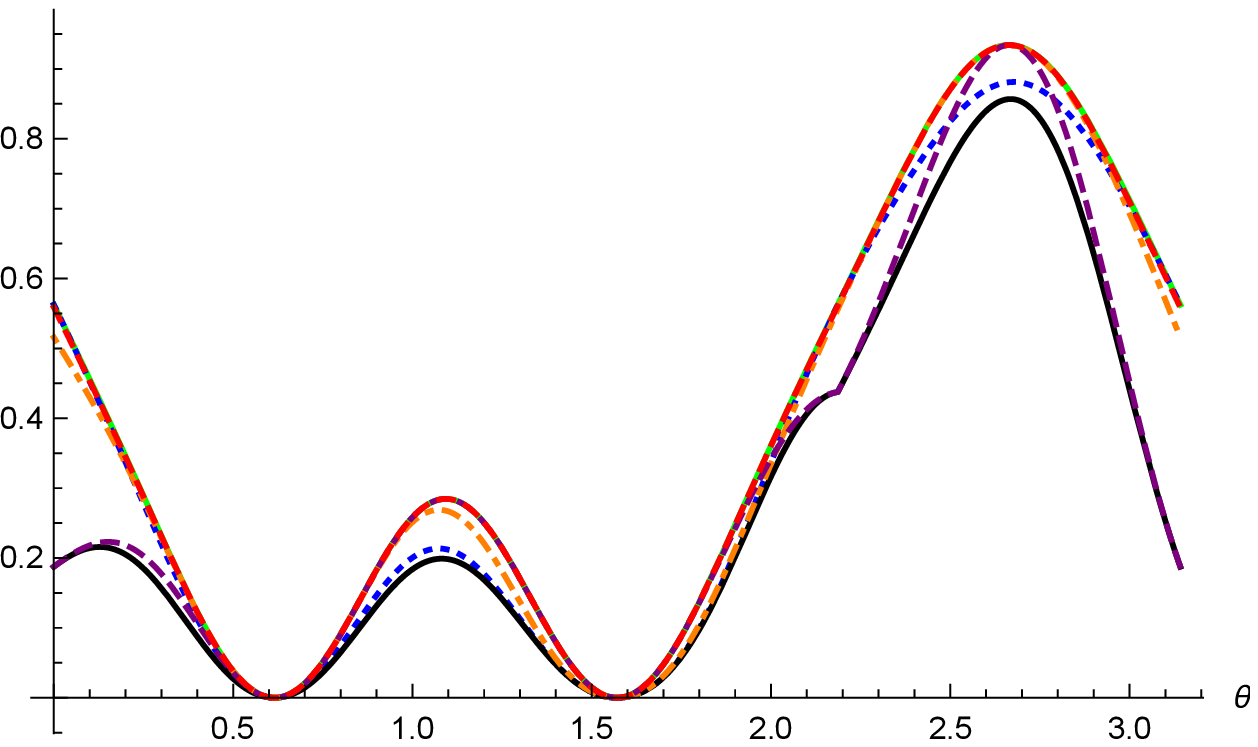}
  \includegraphics[width=6.5cm,height=4cm]{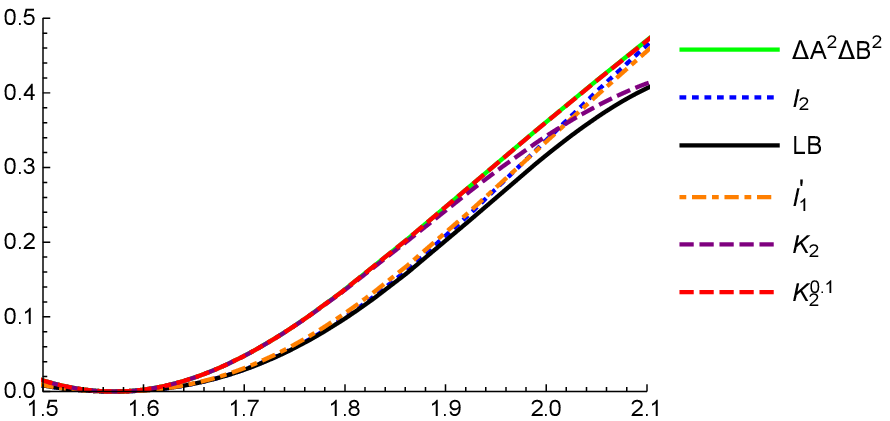}
\caption{\textbf{Comparison of bounds III.} The upper solid green curve is $\Delta A^2\Delta B^2$ and the lower black curve is Bong et al's bound $LB$.
  The blue dotted curve is Yu et al's bound $I_2$ and the orange dotdashed curve is Li et al's bound $I_1^{'}$. Our bounds $K_2$ and $K^{0.1}_2$ are shown in purple dashed and red dashed curves respectively.}
\end{figure}

\begin{figure}[!ht]
\centering
\includegraphics[width=8cm,height=4cm]{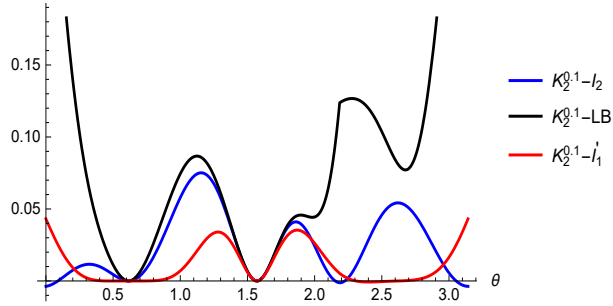}\\
\caption{\textbf{Bound Differences III.} The solid black, blue and red curves represent $K_2^{0.1}-LB$, $K_2^{0.1}-I_2$ and $K_2^{0.1}-I_1^{'}$ respectively. The top black curve reveals that our bound is the best one.}
\end{figure}

\emph{Example} 4. For any two operators $A, B$ in $SU(2)$ and the qubit state $\rho=\frac{1}{2}(I+\vec{r}\cdot \vec{\sigma})$ with $\vec{r}=(r_1,r_2,r_3)\in R^{3}$ and Pauli matrices $\vec{\sigma}=(\sigma_1,\sigma_2,\sigma_3)$.

 We known that $\rho$ is a pure state when $|r|=\sqrt{\sum_{i=1}^3 r^2}=1$, in this case all unitary uncertainty bounds we mentioned above are equal. When $|r|<1$, $\rho$ is a mixed state, we consider the purification of $\rho$ as following. Set $\lambda_i(i=1,2)$ to be the eigenvalues of $\rho$ and $p_i$ to be the corresponding norm eigenvectors. It is known that $\rho$ has the unitary diagonalizable form $\rho=UDU^{\dag}$, where $U=(p_1,p_2)$ is a unitary matrix and $D=diag(\lambda_1,\lambda_2)$. In this article, we consider the vectorization of $\sqrt{\rho}=U\sqrt{D}U^{\dag}$ to be
the purification of the mixed state $\rho$. Specifically, the purification of the state $\rho=\frac{1}{2}(I+\vec{r}\cdot \vec{\sigma})$ is
\begin{equation*}
|\sqrt{\rho}\rangle=\begin{bmatrix}\frac{r_3(-\sqrt{1-|r|}+\sqrt{1+|r|})+|r|(\sqrt{1-|r|}+\sqrt{1+|r|})}{2\sqrt{2}|r|}\\
\frac{(|r|^2-r_3^2)(\sqrt{1-|r|}-\sqrt{1+|r|})}{2\sqrt{2}|r|(r_1+ir_2)}\\
\frac{-(r_1+ir_2)(\sqrt{1-|r|}-\sqrt{1+|r|})}{2\sqrt{2}|r|}\\
\frac{r_3(\sqrt{1-|r|}-\sqrt{1+|r|})+|r|(\sqrt{1-|r|}+\sqrt{1+|r|})}{2\sqrt{2}|r|}\\\end{bmatrix}.
\end{equation*}

Since $SU(2)$ is homeomorphic to the unit sphere, hence a general 2-dimensional unitary operator $A=aI +\sqrt{-1}\sum_{i=1}^3a_i\sigma_i$, where $a, a_i\in \mathbb{R}$ are on the unit sphere: $a^2+\sum_{i=1}^3a_i^2=1$. The modulus of the four coordinates of the vector $v=I\otimes \delta A |\sqrt{\rho}\rangle$ are our real numbers $x_i (i=1,\cdots,4)$. Following our construction in \eqref{eq6}, we can get the closed form of $K_m$ for any two unitary operators and a qubit sate.

\begin{figure}[!ht]
\centering
\includegraphics[width=9cm,height=4cm]{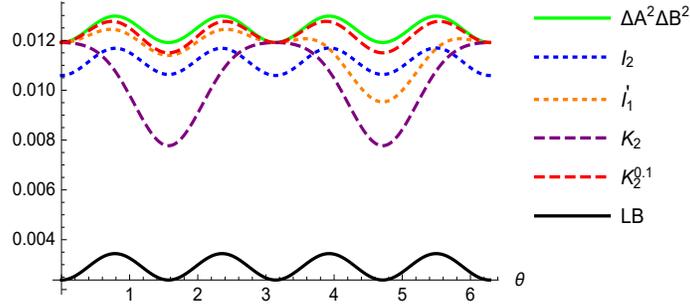}
\caption{\textbf{Comparison of bounds for purification of the state.} The solid green (upper), solid black (lower), dotted blue and orange  curves are $\Delta A^2\Delta B^2$, Bong et al's bound LB,
Yu et al's bound $I_2$ and Li et al's bound $I_1^{'}$ respectively. Our bounds $K_2$ and $K^{0.1}_2$ are shown in purple and red dashed curves respectively.} \label{gra3}
\end{figure}

For example, when $A=cos\frac{\pi}{8}I-\sqrt{-1}sin\frac{\pi}{8}\sigma_2$ and $B=cos\frac{\pi}{8}I+\sqrt{-1}sin\frac{\pi}{8}\sigma_3$
and consider the state $\rho$ with $\vec{r}=(\frac{1}{3},\frac{2}{3}cos\theta, \frac{2}{3}sin\theta)$. After the purification
the bounds are computed according to Theorem 1 (\ref{t:bound}). It turns out that
$\Delta A^2\Delta B^2>K^{0.1}_2>I_2>I_3>I_4\geq LB= |\langle A^{\dagger}B\rangle-\langle A^{\dagger}\rangle\langle B\rangle|^2$.
Figure \ref{gra3} shows the comparison. From the picture it is clear
that the bound $K^{0.1}_2$ outperforms all the bounds $I_2$ from \cite{Yu} and $LB$
from \cite{Bong}, meanwhile the bound $K_2$ is strictly better than Bong's bound $LB$ in whole range.

\begin{figure}[!ht]
\centering
\includegraphics[width=8cm,height=4cm]{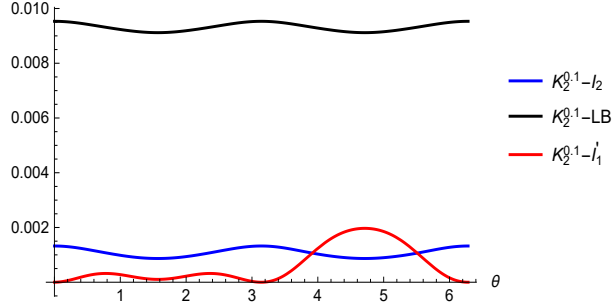}
\caption{\textbf{Bound differences IV.} The solid black, blue and red curves represent $K_2^{0.1}-LB$, $K_2^{0.1}-I_2$ and $K_2^{0.1}-I_1^{'}$ respectively.}\label{gra4}
\end{figure}
\begin{figure}[!ht]
\centering
\includegraphics[width=8cm,height=4cm]{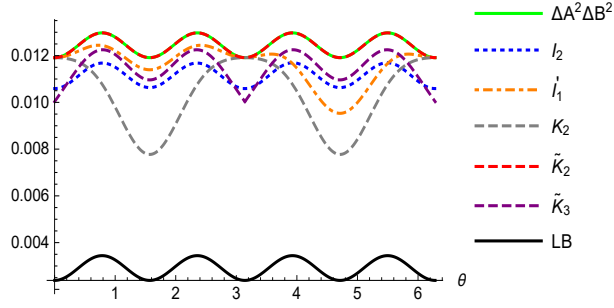}
\caption{\textbf{Bound strengthened by permutation.} The solid green (upper), solid black (lower), dotted blue and dotted dashed orange curves are $\Delta A^2\Delta B^2$, Bong et al's bound LB,
Yu et al's bound $I_2$ and Li et al's bound $I_1^{'}$ respectively. Our bounds $K_2$, $\tilde{K}_2$ and $\tilde{K}_3$ are shown in gray, red and purple dashed curves respectively.}\label{gra5}
\end{figure}

The bound $K_2$ can be further strengthened by Theorem 2 (\ref{t:bound2}). Take the permutation $\sigma=(2~4)$, then  $\tilde{K}=\tilde{K}_2=(2~4)K_2=\Delta A^2\Delta B^2$. Figure \ref{gra4} shows the difference between our bounds and the former bounds. Figure \ref{gra5} depicts the situation, where the curve corresponding to $\tilde{K}_3$ is almost better than that to $I_2$.

\section{Uncertainty relations for multiple unitary operators}

We can generalize the lower bounds to three and more unitary operators easily.
Let $A$, $B$ and $C$ be three unitary operators defined on the Hilbert space $H$
containing a fixed state $\psi$.
By Theorem 1 (\ref{t:bound}) the UURs for the pairs $\{A,B\}$, $\{B,C\}$ and $\{A,C\}$
are as follows.
\begin{align*}
\Delta A^2\Delta B^2\geq K_m, \Delta B^2\Delta C^2\geq K'_m,
\Delta A^2\Delta C^2\geq K_m^{''},
\end{align*}
where $(0\leq m\leq \dim H/2)$.
Taking the square root of the product, we have the following result.

\emph{Proposition} 1. Let $\psi$ be a fixed state in an $n$-dimensional Hilbert space $H$, and $A$, $B$ and $C$ three unitary operators on $H$.
 Then the product of the variances satisfies ($1\leq m\leq [\frac n2]$)
\begin{equation}\begin{split}
\Delta A^2\Delta B^2\Delta C^2&\geqslant (K_m^v(K_m^v)'(K_m^v)'')^{1/2}\geqslant (K_mK'_mK''_m)^{1/2}
\end{split}\end{equation}
where $K_m, K'_m, K''_m$ and  $K_m^v, (K_m^v)', (K_m^v)''$ are the quantities defined in \eqref{t:bound} for the pairs
$\{A, B\}$, $\{A, C\}$ and $\{B, C\}$ respectively.

The lower bound can be tightened by employing the symmetry
given in Theorem 2.

\emph{Proposition} 2. Let $\psi, A, B, C$ be as in Prop. 1. Then the UURs are given by ($1\leq m\leq [n/2]$)
\begin{equation}
\Delta A^2\Delta B^2\Delta C^2\geq (\tilde{K}_m\tilde{K}'_m\tilde{K}''_m)^{1/2},
\end{equation}
where $\tilde{K}_m, \tilde{K}'_m, \tilde{K}''_m$ are the improved
lower bounds \eqref{e:bound2}
for the pairs $\{A, B\}$, $\{A, C\}$ and $\{B, C\}$ respectively. 

The lower bounds can be generalized to multi-observables.

\emph{Theorem 3}. Let
$A_1, \dots, A_l$ be unitary operators on an $n$-dimensional Hilbert space with a
fixed state $\psi$, then for each $1\leq m\leq [\frac n2]$,
\begin{equation}
\Delta^2(A_1)\cdots \Delta^2(A_l)\geq \left(\prod_{1\leq i<j\leq l} K^{ij}_m\right)^{\frac 1{l-1}}
\end{equation}
where $K^{ij}_m=K_m$ given in Theorem 1 for the pair $\{A_i, A_j\}$.
The same inequality also holds for $K^{ij}_m=\tilde{K}_m$ defined in Theorem 2 (\ref{t:bound2})
for the pair $\{A_i, A_j\}$. 

For three unitary operators, the UUR by Bong et al's can be expressed as follows \cite{Bong}:
\begin{equation}
det\begin{bmatrix} 1&\lan A\ra &\lan B\ra &\lan C\ra \\\lan A^{\dagger}\ra &1&\lan  A^{\dagger}B\ra &\lan  A^{\dagger}C\ra\\
\lan B^{\dagger}\ra&\lan  B^{\dagger}A\ra&1&\lan  B^{\dagger}C\ra\\\lan C^{\dagger}\ra&\lan C^{\dagger}A\ra&\lan C^{\dagger}B\ra&1\end{bmatrix}\geq 0,
\end{equation}
which can be rewritten as
\begin{equation}
\begin{split}
\Delta A^2\Delta B^2\Delta C^2\geq& \Delta A^2|\lan B^{\dagger}C\ra-\lan B^{\dagger}\ra\lan C\ra|^2\\
+&\Delta B^2|\lan A^{\dagger}C\ra-\lan A^{\dagger}\ra\lan C\ra|^2
+\Delta C^2|\lan A^{\dagger}B\ra-\lan A^{\dagger}\ra\lan B\ra|^2\\
-&2Re\{(\lan A^{\dagger}C\ra-\lan A^{\dagger}\ra\lan C\ra)\cdot(\lan C^{\dagger}B\ra-\lan C^{\dagger}\ra\lan B\ra)
(\lan B^{\dagger}A\ra-\lan B^{\dagger}\ra\lan A\ra)\}.
\end{split}\end{equation}
In \cite{JSL}, the quantities $M_{tpq}^z, M_{tpq}^x, M_{tpq}^y$ were constructed and shown that $\Delta A^2 \Delta B^2 \Delta C^2 \geq \max\{M_{tpq}^z, M_{tpq}^x, M_{tpq}^y\}$ \cite[Sect.3]{JSL}. Let's consider how our bounds behave in the multi-operator case.

\begin{figure}[!ht]
\centering
\includegraphics[width=10cm,height=4.5cm]{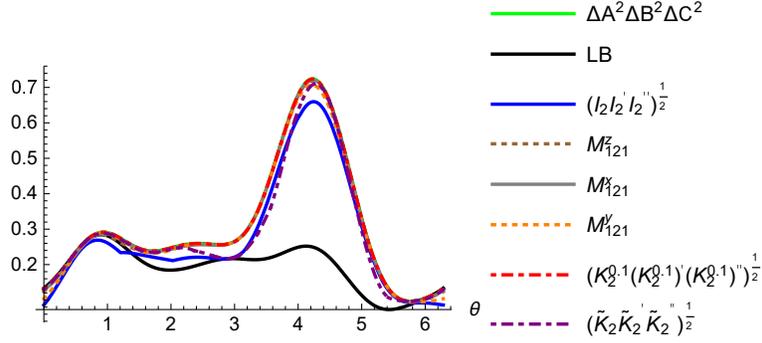}
\caption{\textbf{Comparison of bounds for three operators.} { The solid green (upper), black (lower), blue curves are $\Delta A^2\Delta B^2\Delta C^2$, Bong et al's bound LB and Yu et al's bound respectively. The dotted grown, solid gray and dotted orange curves are Li et al's bounds $M^{z}_{121}$, $M^{x}_{121}, M^{y}_{121}$ respectively. The red and purple dotted dashed curves represent our improved bounds $(K_2(K_2^{0.1})'(K_2^{0.1})'')^{\frac{1}{2}}$ and $(\tilde{K}_2\tilde{K}'_2\tilde{K}''_2)^{\frac{1}{2}}$ respectively.} }\label{grap10}
\end{figure}

\begin{figure}[!ht]
\centering
\includegraphics[width=6.5cm,height=4cm]{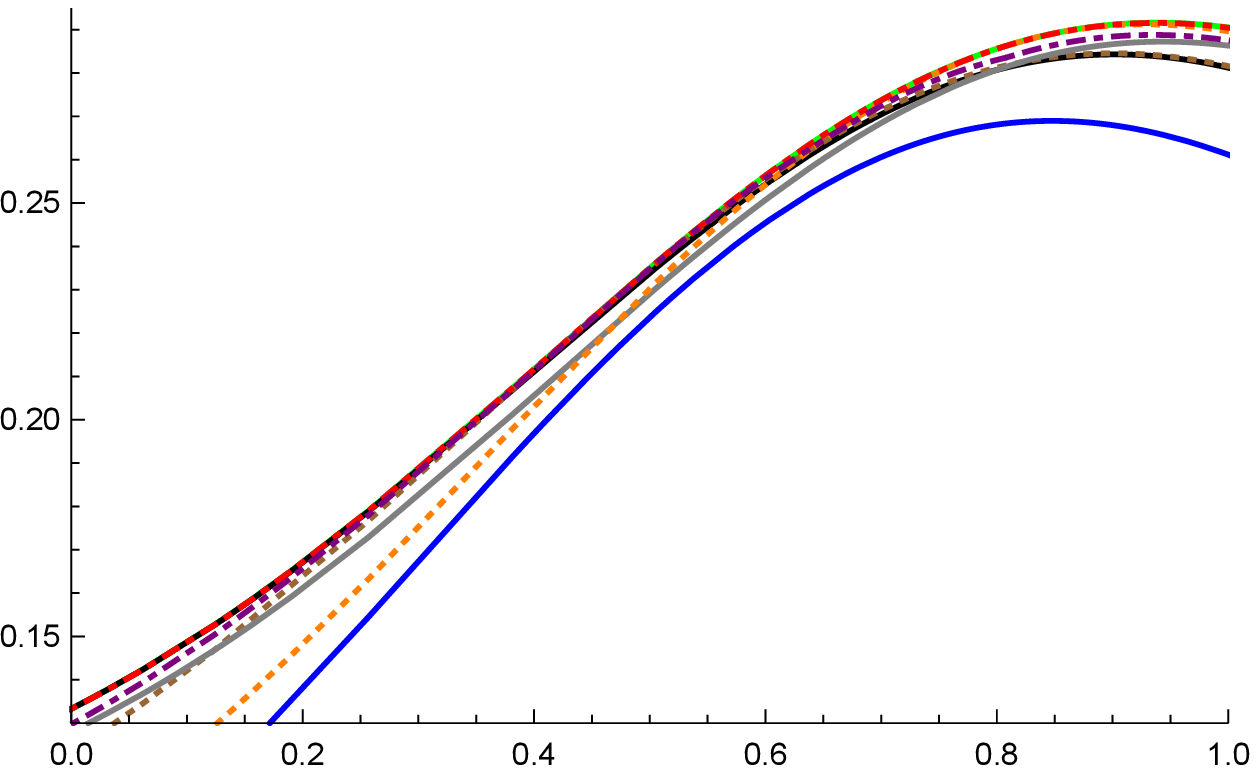}
\includegraphics[width=6.5cm,height=4cm]{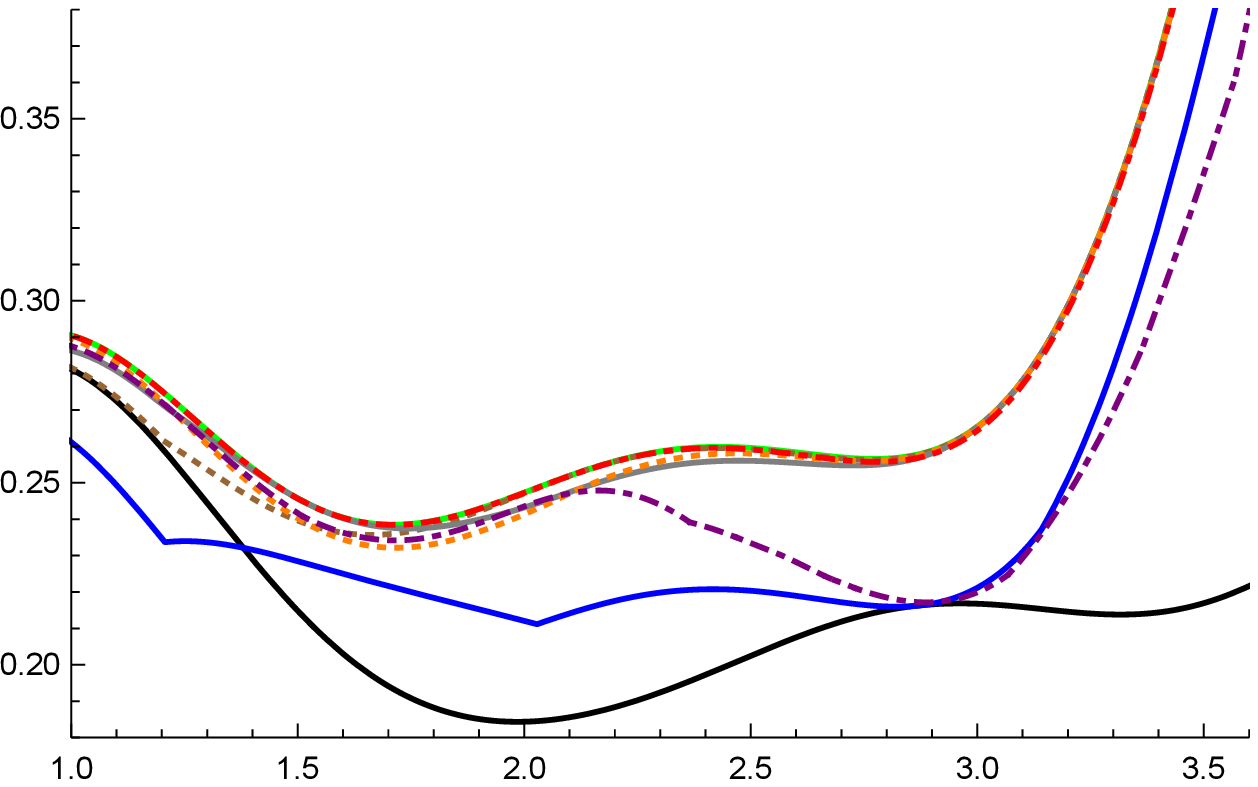}\\
\includegraphics[width=6.5cm,height=4cm]{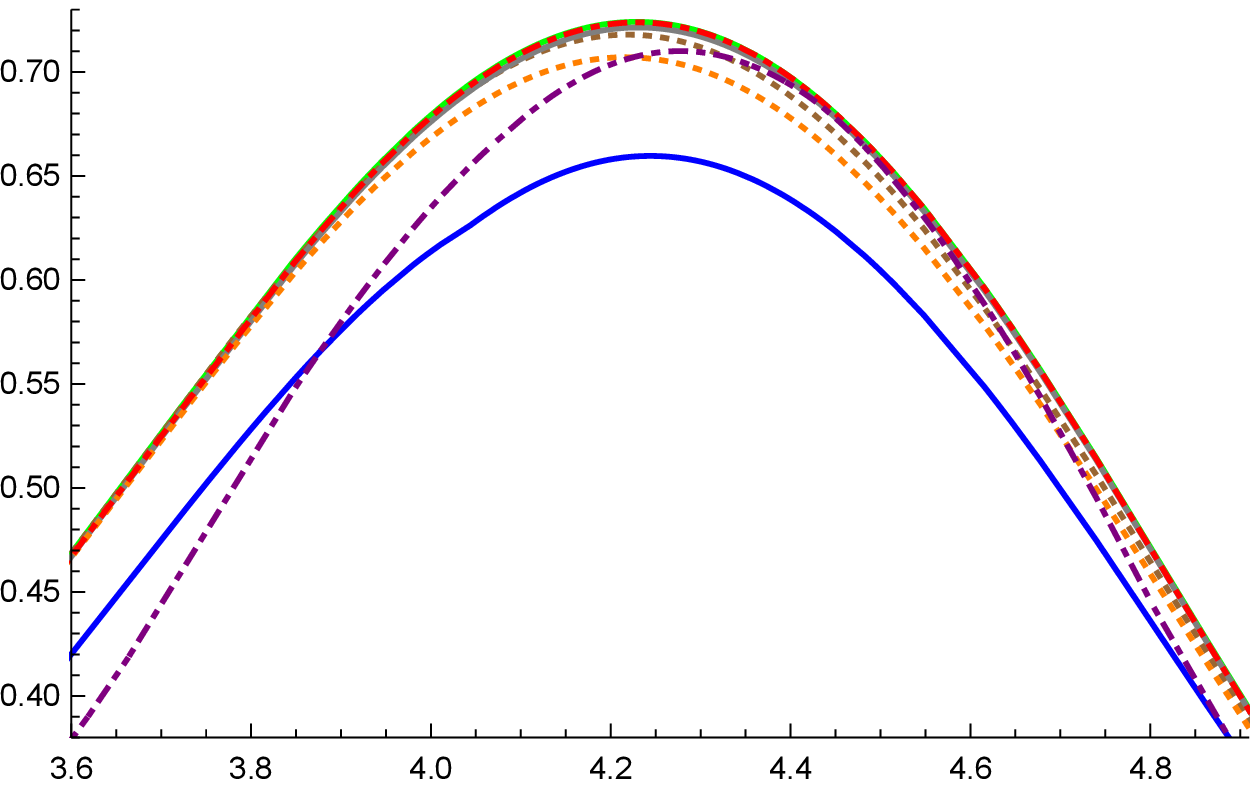}
\includegraphics[width=6.5cm,height=4cm]{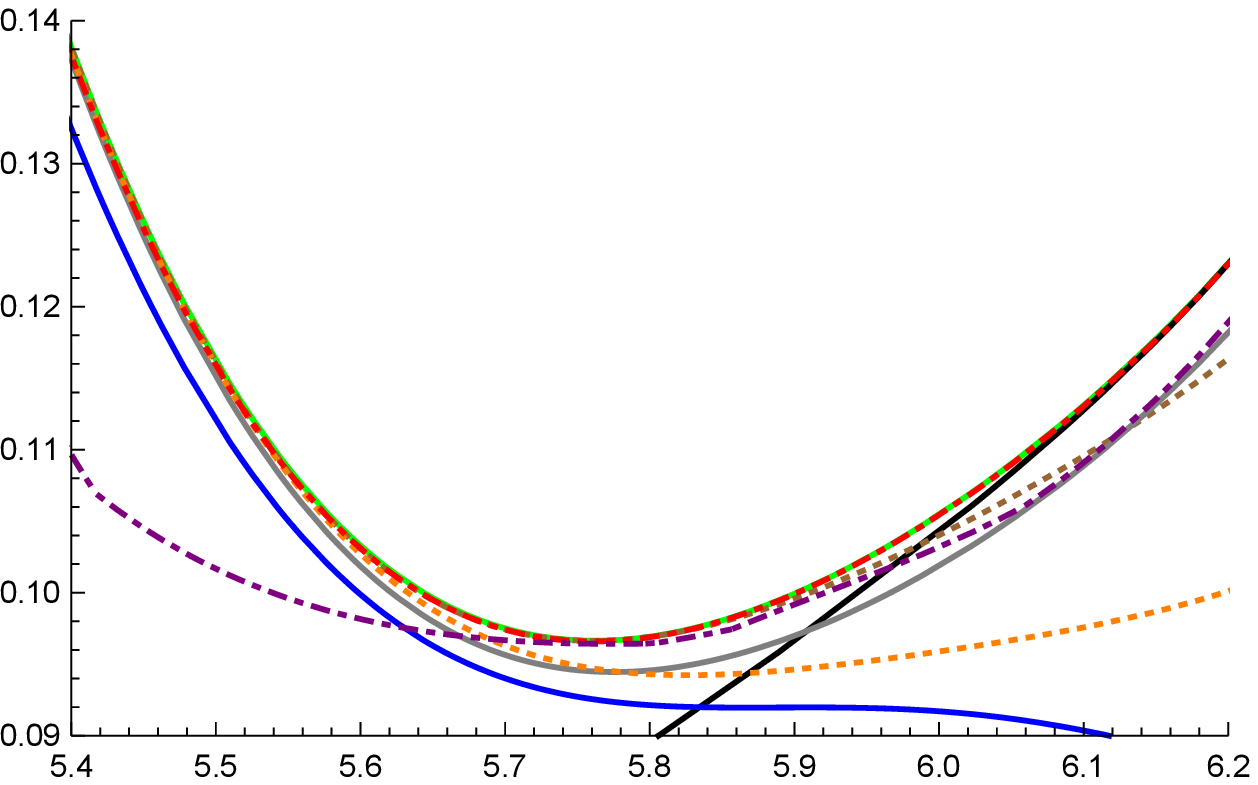}
\caption{\textbf{Zoom-in of Figure 10.} The red dotted dashed curve (corresponding to ours) stand out among all the bounds!}\label{grap11}
\end{figure}

\begin{figure}[!ht]
\centering
\includegraphics[width=6cm,height=4cm]{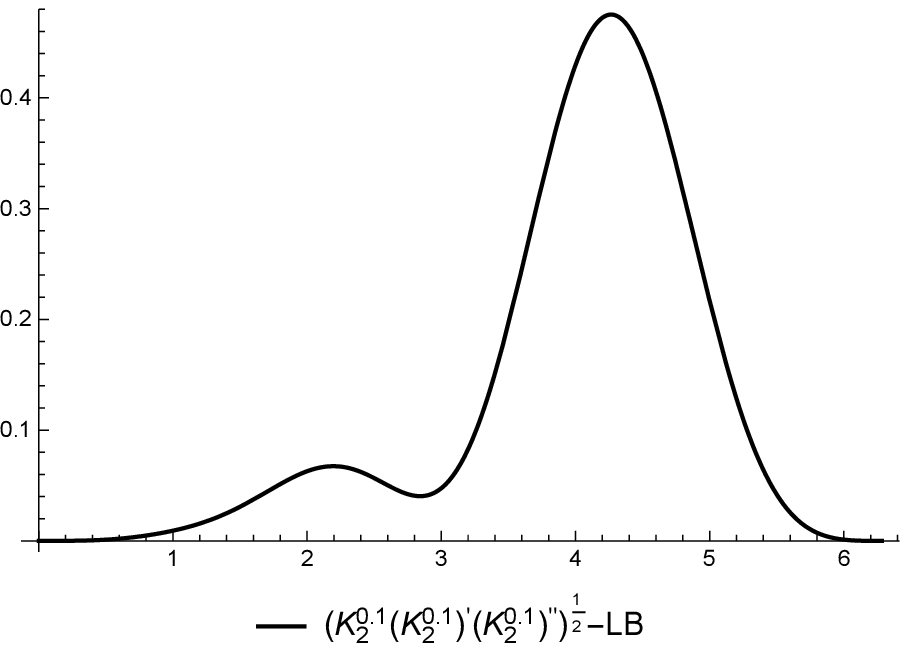}
\includegraphics[width=8cm,height=3.5cm]{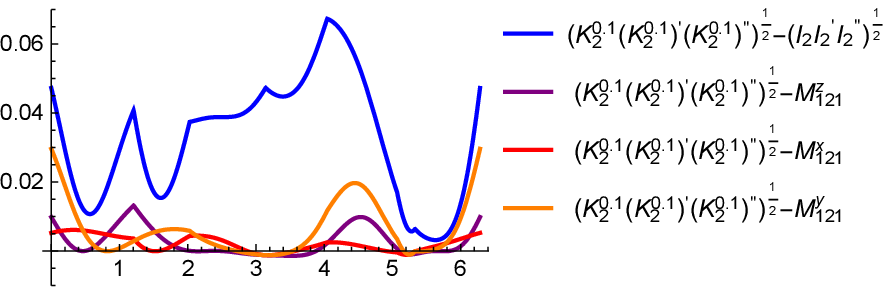}
\caption{\textbf{Bound difference V.}  The solid black curve represents
 $(K_2^{0.1}(K_2^{0.1})'(K_2^{0.1}))^{\frac{1}{2}}-LB$. The solid blue, purple, red, and orange curves are the differences between
$(K_2^{0.1}(K_2^{0.1})'(K_2^{0.1}))^{\frac{1}{2}}$ and $(I_2I_2^{'}I_2^{''})^{\frac{1}{2}}$, $M^{z}_{121}$, $M^{x}_{121}$ and $M^{y}_{121}$
respectively.}\label{grap12}
\end{figure}

\emph{Example} 5. Let $|\psi\ra=\frac{1}{2}cos\frac{\theta}{2}|0\ra+\frac{\sqrt{3}}{2}sin\frac{\theta}{2}|1\ra+\frac{1}{2}sin\frac{\theta}{2}|2\ra
+\frac{\sqrt{3}}{2}cos\frac{\theta}{2}|3\ra$ and consider three unitary operators
\begin{equation*}
A=\begin{bmatrix} 1&0&0&0\\0&i&0&0\\0&0&-1&0\\0&0&0&-i\end{bmatrix},
 B=\begin{bmatrix} 0&1&0&0\\1&0&0&0\\0&1&0&0\\0&0&1&0  \end{bmatrix},
 C=\begin{bmatrix} 0&1&0&0\\1&0&0&0\\0&0&1&0\\0&0&0&-1\end{bmatrix}.
\end{equation*}
The lower bounds $(K_2(K_2^{0.1})'(K_2^{0.1})'')^{\frac{1}{2}}$ and $(\tilde{K}_2\tilde{K}'_2\tilde{K}''_2)^{\frac{1}{2}}$,
calculated using Prop. 1 and  Prop. 2, are better
than those of Yu et al's and Bong et al.'s over a significant region. See Figures \ref{grap10}, \ref{grap11} and \ref{grap12} for the comparison, where the latter two charts highlight the differences.

\begin{figure}[!ht]
\centering
\includegraphics[width=7cm,height=6cm]{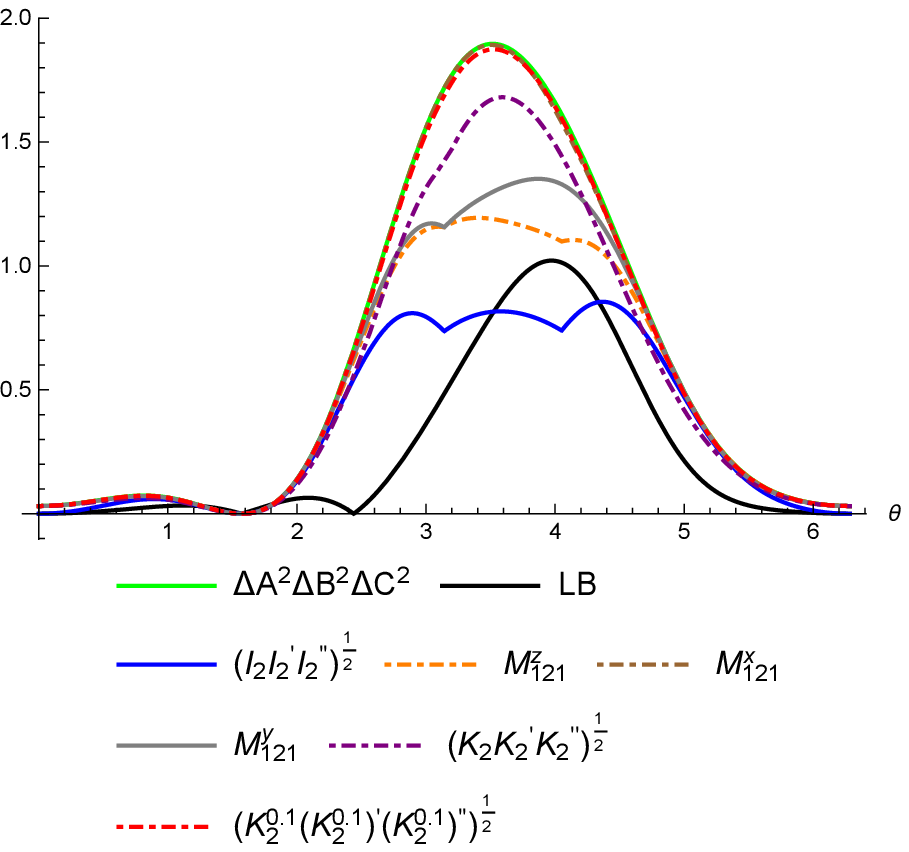}
\includegraphics[width=6cm,height=5cm]{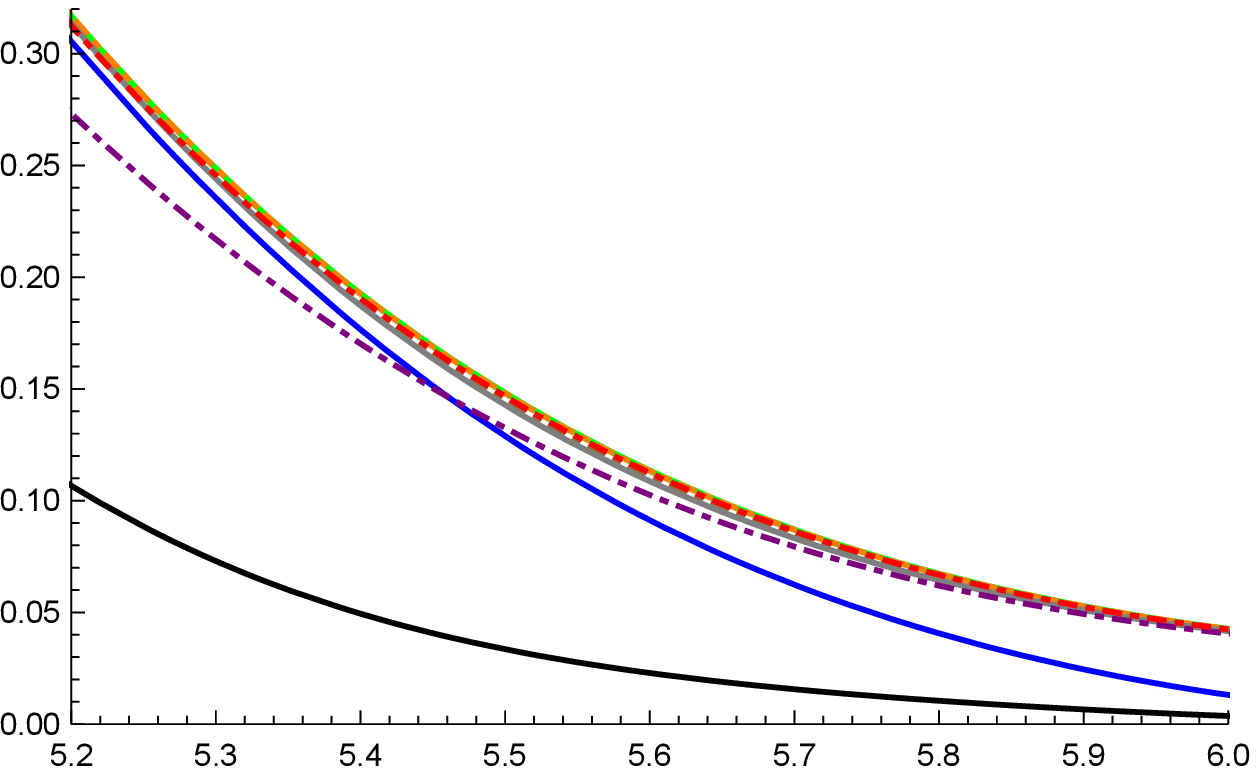}
\caption{\textbf{Comparison for three operators.} The solid green (upper), black (lower) and blue curves are $\Delta A^2\Delta B^2\Delta C^2$, Bong et al's bound LB and the Yu et al's best bound. The solid gray, dotted dashed brown, orange curves are Li et al's bounds $M^{z}_{121}$, $M^{x}_{121}, M^{y}_{121}$ respectively. The red and purple dotted dashed curves represent our bounds $(K_2K'_2K''_2)^{\frac{1}{2}}$ and $(K_2^{0.1}(K_2^{0.1})'(K_2^{0.1})'')^{\frac{1}{2}}$ respectively.}\label{grap13}
\end{figure}

\begin{figure}[!ht]
\centering
\includegraphics[width=12cm,height=4.5cm]{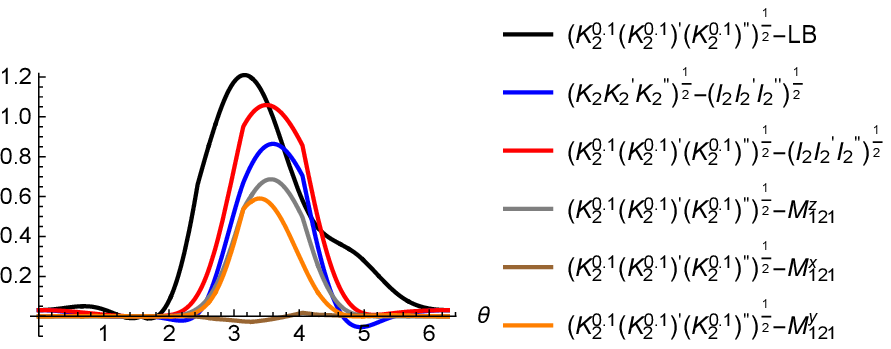}
\caption{\textbf{Bound differences VI.} The solid black, blue and red curves represent the bound differences
$(K_2^{0.1}(K_2^{0.1})'(K_2^{0.1})^{''})^{\frac{1}{2}}-LB$, $(K_2K_2^{'}K_2^{''})^{\frac{1}{2}}-(I_2I_2^{'}I_2^{''})^{\frac{1}{2}}$ and $(K_2^{0.1}(K_2^{0.1})^{'}(K_2^{0.1})^{''})^{\frac{1}{2}}-(I_2I_2^{'}I_2^{''})^{\frac{1}{2}}$ respectively. The solid gray, brown and orange curves are the difference between $(K_2^{0.1}(K_2^{0.1})'(K_2^{0.1}))^{\frac{1}{2}}$ with $M^{z}_{121}$,$M^{x}_{121}$, $M^{y}_{121}$ respectively.}\label{grap14}
\end{figure}

\emph{Example} 6. For a 3-dimensional Hilbert space, let us consider the pure state $\frac{\sqrt{2}}{2}cos\frac{\theta}{2}|0\ra+\frac{\sqrt{2}}{2}sin\frac{\theta}{2}|1\ra-sin\frac{\theta}{2}|2\ra$.
 Suppose the three unitary operators are as follows:
\begin{equation*}
\begin{split}
A=\begin{bmatrix} 1&0&0\\0&e^{\frac{i\pi}{2}}&0\\0&0&e^{\frac{3i\pi}{2}}\end{bmatrix},
 B=\begin{bmatrix} 0&0&1\\1&0&0\\0&1&0 \end{bmatrix},
 C=\begin{bmatrix} 0&1&0\\1&0&0\\0&0&1\end{bmatrix}.
  \end{split}
\end{equation*}
The lower bounds $(K_2K'_2K''_2)^{\frac{1}{2}}$ and $(K_2(K_2^{0.1})'(K_2^{0.1})'')^{\frac{1}{2}}$  can be calculated using the Propositions. We see that the lower bound $(K_2(K_2^{0.1})'(K_2^{0.1})'')^{\frac{1}{2}}$ is mostly tighter than those of Yu et al's, Bong et al's and Li et al's. See Figures \ref{grap13} and \ref{grap14} for the comparison.

All these examples have shown that our new bounds are mostly tighter than the previously known bounds in the case of multi-operators as well.

\section{\label{sec:leve4} Conclusion}
We have proposed an improved product-form of variance-based unitary uncertainty relations (UUR)
for unitary operators relative to pure quantum states. The celebrated Heisenberg-Robertson's uncertainty relation reveals the
distinguished property of the quantum theory, and is usually formulated with the help of the Cauchy-Schwarz inequality.
Our approach to the problem is to get to the bottom of the inequality to improve the
lower bounds for the UURs by using brackets and convex functions.

We have shown that the new uncertainty bounds outperform almost all existing UURs in the literature,
notably better than Bong et al's strong UUR using the positivity of the Gram matrix \cite{Bong} for two and certain
multiple unitary operators. This is achieved by mathematical analysis and
illustrated by six examples
where our new bounds are simpler and tighter than both lower bounds of the UURs given in \cite{Bong, Yu, JSL}. The new bounds may shed new light on the effect of
fine-grained inequalities and also the process how the data are collected
in the measurement. We have also analyzed the effect of mixed states and pointed out that
only UURs for pure states are enough for theoretical foundation of uncertainty relations in the product-form.

\bigskip

\centerline{\bf Acknowledgments}
The research is supported in part by the NSFC
grant nos.
11871325, 12126351 and 12126314, and the Natural Science Foundation of Hubei Province
grant no. 2020CFB538 as well as the Simons Foundation
grant no. 523868.
 \vskip 0.1in

\bibliographystyle{amsalpha}

\begin{thebibliography}{ABCD}
\bibitem{1} Fuchs C.A. and Peres A.: Quantum-state disturbance versus information gain: Uncertainty relations for quantum information. Phys. Rev. A $\mathbf{53}$, 2038 (1996).

\bibitem{2} Renes J.M. and Boileau J.-C.: Conjectured strong complementary information tradeoff. Phys. Rev. Lett. $\mathbf{103}$, 020402 (2009).

\bibitem{3} Bowen W.P., Schnabel R., Lam P.K. and Ralph T.C.: Experimental investigation of criteria for continuous variable entanglement.
Phys. Rev. Lett. $\mathbf{90}$, 043601 (2003). 

\bibitem{4} G\"uhne O.: Characterizing entanglement via uncertainty relations. Phys. Rev. Lett. $\mathbf{92}$, 117903 (2004).

\bibitem{5} Howell J.C., Bennink R.S., Bentley S.J. and Boyd R.W.: Realization of the Einstein-Podolsky-Rosen Paradox Using Momentum- and Position-Entangled Photons from Spontaneous Parametric Down Conversion. Phys. Rev. Lett. $\mathbf{92}$, 210403 (2004).

\bibitem{6} Pires D.P., Cianciaruso M., C\'eleri L.C., Adesso G. and Soares-Pinto D.O.: Generalized geometric quantum speed limits. Phys. Rev. X $\mathbf{6}$, 021031(2016).

\bibitem{7}  Cand\'es E.J., Romberg J. and Tao T.:Robust uncertainty principles: Exact signal reconstruction from highly incomplete frequency information. IEEE. Trans. Inf. Theory $\mathbf{52}$, 489(2006).

\bibitem{8} Erhart J., Sponar S., Sulyok G., Badurek G., Ozawa M. and Hasegawa Y.: Experimental demonstration of a
universally valid error-disturbance uncertainty relation in spin measurements. Nat. Phys. $\mathbf{8}$, 185 (2012).

\bibitem{9} Sulyok G., Sponar S., Demirel B., Buscemi F., Hall M.J.W., Ozawa M. and Hasegawa Y.: Experimental test of entropic noise-disturbance uncertainty relations for spin-1/2 measurements. Phys. Rev. Lett. $\mathbf{115}$, 030401 (2015).

\bibitem{10} Li C.F., Xu J.S., Xu X.Y., Li K. and Guo G.C.: Experimental investigation of the entanglement-assisted entropicuncertainty principle. Nat. Phys. $\mathbf{7}$, 752 (2011).

\bibitem{11} Rozema L.A., Darabi A., Mahler D.H., Hayat A., Soudagar Y. and Steinberg A.M.: Violation of Heisenbergs measurement-disturbance relationship by weak measurements. Phys. Rev. Lett. $\mathbf{109}$, 100404 (2012).

\bibitem{12} Weston M.M., Hall M.J.W., Palsson M.S., Wiseman H.M. and Pryde G.J.: Experimental test of universal complementarity relations. Phys. Rev. Lett. $\mathbf{110}$, 220402 (2013).

\bibitem{13} Kaneda F., Baek S.-Y., Ozawa M. and Edamatsu K.: Experimental test of error-disturbance uncertainty relations by weak measurement. Phys. Rev. Lett. $\mathbf{112}$, 020402 (2014). 

\bibitem{15} Heisenberg W.: \"uber den anschaulichen Inhalt der quantentheoretischen Kinematik und Mechanik. Z. Phy. $\mathbf{43}$, 172 (1927).

\bibitem{16} Kennard E.H.:Zur quantenmechanik einfacher bewegungstypen. Z. Phys. $\mathbf{44}$, 4 (1927).

\bibitem{17} Weyl H.: Gruppentheorie und Quantenmechanik, Hirzel, Leipzig (1928)

\bibitem{18} Robertson H.P.: The uncertainty principle. Phys. Rev. $\mathbf{34}$, 163 (1929).

\bibitem{19} Colangelo G., Ciurana F.M., Bianchet L.C., Sewell R.J. and Mitchell M.W.: Simultaneous tracking of spin angle and amplitude beyond classical limits. Nature $\mathbf{543}$, 525(2017). 

\bibitem{20} Schwonnek R., Reeb D. and Werner R.F.: Measurement uncertainty for finite quantum observables. Mathematics, {\bf4}, 38 (2016).

\bibitem{21} Bialynicki-Birula I. and Mycielski J.: Uncertainty relations for information entropy in wave mechanics. Commun. Math. Phys. {\bf44}, 129 (1975).

\bibitem{22} Deutsch D.: Uncertainty in quantum measurements. Phys. Rev. Lett. {\bf50}, 631 (1983). 

\bibitem{23} Maassen H. and Uffink J.B.M.: Generalized entropic uncertainty relations. Phys. Rev. Lett. {\bf60}, 1103 (1988). 

\bibitem{24} Wehner S. and Winter A.: Entropic uncertainty relations, a survey. New J. Phys. {\bf12}, 025009 (2010).

\bibitem{25} Coles P.J. and Piani M.: Improved entropic uncertainty relations and information exclusion relations. Phys.Rev. A {\bf89}, 022112 (2014).

\bibitem{26} Coles P.J., Berta M., Tomamichel M. and Wehner S.: Entropic uncertainty relations and their applications. Rev. Mod. Phys. {\bf89}, 015002 (2017).

\bibitem{XJ2} Xiao Y., Jing N., Li-Jost X. and Fei S.M.: Improved uncertainty relation in the presence of quantum memory. J. Phys. A: Math. Theor. {\bf49} (2016) 49LT01 (9pp).

\bibitem{MP} Maccone L. and Pati A.K.: Stronger uncertainty relation for all incompatible observables. Phys. Rev. Lett. {\bf113}, 260401 (2014).

\bibitem{XJ1} Xiao Y., Jing N., Li-Jost X. and Fei S.M.: Weighted Uncertainty Relations. Sci. Rep. {\bf6}, 23201 (2016). 

\bibitem{BP} Bagachi S. and Pati A.K.: Uncertainty relations for general unitary operators. Phys. Rev. A {\bf94}, 042104 (2016). 

\bibitem{Busch} Busch P., Lahti P. and Werner R.F.: Colloquium: Quantum root-mean-square error and measurement uncertainty relations. Rev. Mod. Phys. $\bf{86}$, 1261 (2014).

\bibitem{Xiao2019} Xiao Y., Guo C., Meng F., Jing N. and Yung M.-H.: Incompatibility of observables as state-independent bound of uncertainty relations. Phys. Rev. A {\bf100}, 032118 (2019).

\bibitem{Ozawa2003} Ozawa M.: Universally valid reformulation of the Heisenberg uncertainty principle on noise and disturbance in measurement. Phys. Rev. A {\bf67}, 042105 (2003).

\bibitem{SP} Massar S. and Spindel P.: Uncertainty Relation for the Discrete Fourier Transform. Phys. Rev. Lett. $\bf{100}$, 190401 (2008).

\bibitem{Rud} Rudnicki L., Tasca D.S. and Walborn S.P.: Uncertainty relations for characteristic functions. Phys. Rev. A $\bf{93}$, 022109 (2016).

\bibitem{Bong} Bong K.-W., Tischler N., Patel R.B., Wollmann S., Pryde G.J. and Hall M.J.W.: Strong Unitary and Overlap Uncertainty Relations: Theory and Experiment. Phys. Rev. Lett. $\bf{120}$, 230402 (2018).

\bibitem{H2} Schr\"odinger E.: ''Zum Heisenbergschen Unschfeprinzip'' Sitzungsber. Preuss. Akad. Wiss., Phys. Math. Kl. $\bf{19}$, 296 (1930).

\bibitem{R2} Robertson H.P.: A general formulation of the uncertainty principle and its classical interpretation. Phys. Rev. A $\bf{35}$, 667 (1930); 

\bibitem{Yu} Yu B., Jing N. and Li-Jost X.: Strong unitary uncertainty relations. Phys. Rev. A {\bf100}, 022116 (2019).

\bibitem{JSL} Li J., Zhang S., Liu L. and Bai C.M.: An improved bound for strong unitary uncertainty relations with refined sequence. Laser Phys. Lett. {\bf17}, 015201 (2020).

\bibitem{XJ3} Xiao Y., Jing N., Yu B., Fei S.-M. and Li-Jost X.: Strong variance-based uncertainty relations and uncertainty intervals. arXiv:1610.01692.

\bibitem{Lax} Lax P.D., Linear algebra and its apllication, 2nd Ed.(Wiley, New Jersey, 2007).

\end{thebibliography}

\end{document}